\documentstyle [epsf]{mn}
\def\be{\begin{equation}}
\def\ee{\end{equation}}
\def\ba{\begin{eqnarray}}
\def\ea{\end{eqnarray}}
\def\rms{{\em rms }}

\begin{document}

\title[Survival of substructure]
{\bf Survival of Substructure within Dark Matter Haloes}
\author[G.Tormen, A.Diaferio and D.Syer]
{Giuseppe Tormen, Antonaldo Diaferio and David Syer\\
Max-Planck-Institut f\"{u}r Astrophysik,
Karl-Schwarzschild-Strasse 1, 85740 Garching bei M\"{u}nchen - GERMANY\\
\smallskip
Email: bepi@mpa-garching.mpg.de; diaferio@mpa-garching.mpg.de; 
syer@mpa-garching.mpg.de}

\date{Submitted to MNRAS, December 1997}

\maketitle

\begin{abstract}
Using high resolution cosmological $N$-body simulations, we
investigate the survival of dark matter satellites falling into larger
haloes.  Satellites preserve their identity for some time after
merging.  We compute their loss of mass, energy and angular momentum
as dynamical friction, tidal forces and collisions with other
satellites dissolve them.  We also analyse the evolution of their 
internal structure.  Satellites with less than a few per cent the mass
of the main halo may survive for several billion years, whereas larger
satellites rapidly sink into the center of the main halo potential
well and lose their identity. Penetrating encounters between
satellites are frequent and may lead to significant mass loss and
disruption.  Only a minor fraction of cluster mass (10 per cent on
average) is bound to substructure at most redshifts of interest.  We
discuss the application of these results to the survival and extent of
dark matter haloes associated with cluster galaxies, and to interactions
between galaxies in clusters.  We find that $\approx 35-40$ per cent
of galaxy dark matter haloes are disrupted by the present time.
The fraction of satellites undergoing close encounters is similar 
to the fraction of interacting or merging galaxies in clusters at 
moderate redshift. 

\end{abstract}


\section{Introduction}\label{sec:intro}

Kinematics and dynamics of galaxies within clusters are fundamental 
in understanding galaxy formation and evolution.
In models of hierarchical clustering this issue is closely related
to the collisionless dynamics of dark matter haloes hosting individual
galaxies. The fate of these haloes, after merging with other haloes, 
depends on their relative mass, velocity, orbit parameters, and their 
internal structure in an interconnected way. The complexity of this 
problem has led to different approximations describing the interaction
in terms of distinct physical processes treatable analytically. 
These processes include dynamical friction (e.g. White 1976; Binney \& 
Tremaine 1987), tidal forces (Mamon 1993 and references therein)
and resonant orbit coupling (Weinberg 1994). 

The analytic approach provides a way to study the evolution of 
interacting haloes, and to estimate the survival time of satellites 
falling into the potential well of larger systems (Spitzer 1958;
White \& Rees 1978). These predictions have been tested with 
well-designed numerical models, where all the different parameters 
of the problem are under control; in this case, the accuracy and 
limitations of each approximation can be assessed (e.g. Aguilar \& 
White 1985; Moore, Katz \& Lake 1996). 

However, an overall evaluation of the dynamical evolution and 
survival of haloes in a larger system and in a fully cosmological 
context has been investigated only recently (Klypin, Gottl\"ober 
\& Kravtsov 1997; Ghigna et al. in preparation).
The evolution of clustering in a specific cosmological model 
naturally provides the choice and combination of the different 
parameters: the mass distribution of haloes, their merging rates, 
orbital parameters, internal density and velocity profiles.

The analysis of surviving substructure in a cosmological context
also is of great interest since recent semi-analytical models of 
galaxy formation require a self-consistent recipe for the efficiency 
of galaxy merging (Kauffmann, White \& Guiderdoni 1993; Baugh, Cole 
\& Frenk 1996).

Unfortunately, a cosmological approach to this problem has two main
inconveniences. First, the required numerical resolution, both in 
mass and force, is usually inadequate. Secondly, the ability to follow
in detail the merging history and fate of each satellite requires a non 
negligible amount of work and computer resources. 
This problem becomes more serious for simulations with millions of 
particles, which would be more suited for this study.

In the present paper, we overcome these problems by using high 
resolution $N$-body simulations of individual galaxy clusters 
extracted from lower resolution cosmological simulations, as 
described in Tormen, Bouchet \& White (1997).  
Our simulations have both a manageable size and sufficient resolution 
to address the survival of substructure accreted by galaxy clusters. 
The present investigation complements the work presented in Tormen 
(1997), where we studied the merging history of all the dark matter 
haloes formed in the simulations, and followed the haloes until
they were accreted by the main progenitors of the final clusters.

The plan of the paper is as follows.
Section~\ref{sec:met} briefly describes the simulations and the algorithm 
used to identify dark matter haloes. 
In Section~\ref{sec:dftf} we recall the physical processes of interest
for this problem and show a few examples of evolution of substructure. 
Section~\ref{sec:orb} deals with the evolution of halo orbits inside 
the cluster, and in Section~\ref{sec:surv} we study the global mass loss 
for the accreted satellites, and define the corresponding survival times.
Section~\ref{sec:coll} studies close encounters between satellites 
within the cluster, and estimates their relevance to the disruption
of substructure. 
Section~\ref{sec:instr} describes the evolution of the satellite internal 
structure.
In Section~\ref{sec:msub} we measure the fraction of cluster mass
bound to substructure, and in Section~\ref{sec:disc} we discuss a few 
astrophysical applications of the present results.
We finally summarize our results in Section~\ref{sec:conc}.

\section{Method}\label{sec:met}

\subsection{The simulations}\label{sec:sim}
\begin{figure}
\centering
\epsfxsize=\hsize\epsffile{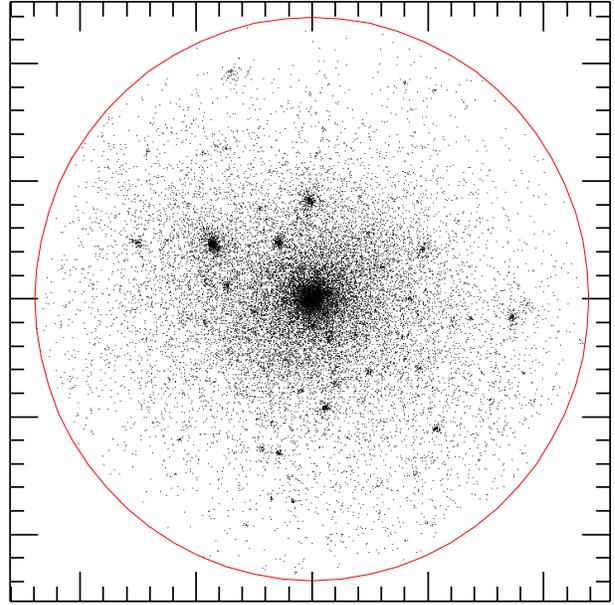}
\caption{One of the nine clusters, at $z=0$. Only particles within the
virial radius are plotted. The tick spacing is 200 kpc.
Notice the amount of resolved substructure.}
\label{fig:g57}
\end{figure}

We use the nine $N$-body simulations described in Tormen et al. (1997).
Each simulation models the formation of an individual galaxy cluster in 
an Einstein--de Sitter universe, with an initial density perturbation 
power spectrum $P(k) \propto k^{-1}$. More massive particles model the
external tidal field (see Tormen et al. 1997 for details).
The Hubble constant is $H_0 = 50$ km s$^{-1}$ Mpc$^{-1}$.
Each final halo is resolved by $\approx 20000$ dark matter
particles within its virial radius $R_v$, each particle with mass 
$m_p \sim 5 \times 10^{10} M_{\sun}$. The effective force 
resolution is $\sim 0.01 R_v$. An example of cluster at $z=0$ is
shown in Figure~\ref{fig:g57}.

The slope of $P(k)$ is chosen to mimic a standard CDM model on 
the scale of galaxy clusters. However, on galactic scales our 
spectrum is shallower than the CDM spectrum, and this results in 
much more small-scale structure than one expects in a corresponding 
CDM simulation with comparable resolution. This feature creates a 
larger statistics for the substructure. As the orbital parameters 
of satellites accreted by a cluster depend very little on the 
cosmology (Tormen, Frenk \& White, in preparation), our results 
will also hold for other cosmological models.

\subsection{Identification of substructure}\label{sec:hid}

We identify dark matter satellites using an overdensity criterion. 
Each satellite groups all the particles within a sphere of mean 
density contrast $\rho/\rho_b = 178$ (with $\rho_b$ the mean
background density), centred on the local minimum of the potential 
energy (see Tormen 1997 for details). 
The corresponding radius of the satellite is its virial radius $r_v$.
 
We wish to study the fate of haloes merging with each other. 
To simplify this task, we focus our study on satellites accreted directly 
onto the main progenitor of the main halo in each simulation.
The main progenitor of a halo is chosen following its merging 
history back in time, and selecting the main branch at each split of 
the merging tree (see e.g. Lacey and Cole 1993).
For each simulation, this procedure provides a catalogue of 
satellites which will merge with the main progenitor halo at different times.

The {\em identification time} $t_{id}$ of a satellite is defined as 
the last time output before the satellite accretes onto the main halo
progenitor. 
The {\em merging time} $t_{mer}$ of a satellite 
is the time when the satellite first crosses the 
virial radius of the major halo. Consecutive outputs of our 
simulations are separated by an interval $\Delta t \simeq 1.6$ Gyr.
Therefore, the merging time for a satellite can only be bracketed between
$t_{id}$ and $t_{id} + \Delta t$. We assign to each satellite
a nominal merging time $t_{mer} = t_{id} + 0.5\Delta t$.

After a satellite is accreted by a larger halo, its orbit and its
internal structure are perturbed by the interaction with the main 
halo and with other substructure, and the satellite may eventually 
dissolve. To keep track of its motion after merging, we consider 
two different masses associated with the satellite: 
{\em(i)} the fraction of its initial mass which remains self-bound
(Section~\ref{sec:sb}), and {\em (ii)} the fraction of its initial 
mass within its tidal radius (Section~\ref{sec:tidal}).
These definitions identify the satellite as a separate 
dynamical entity within the main halo. 

\begin{figure*}
\centering
\epsfxsize=\hsize\epsffile{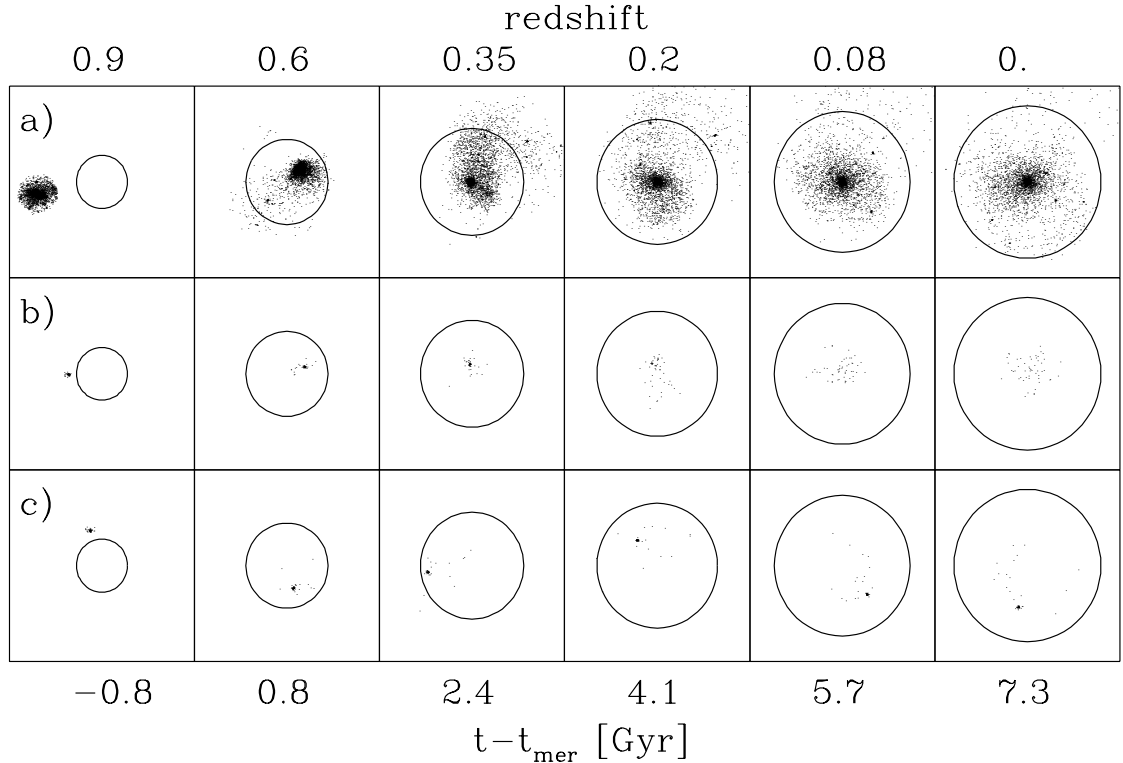}
\caption{Snapshots of three mergers.  The circle denotes the virial
radius of the parent halo.  Only the particles initially within the
virial radius of the satellite are plotted in each frame.
To give an idea of our resolution, the satellite in the third 
example is formed by 68 particles, of which 51 are still self-bound 
in the final output.}
\label{fig:part}
\end{figure*}

The simulations were run for a Hubble time, i.e. $\approx 13$ Gyr for
our cosmological model. Because we consider satellites which merged
between $z\simeq 2.8$ and $z\simeq 0.07$, we can follow them within
the cluster for up to $\sim 11$ Gyr, depending on their merging time.
We limit the present analysis to satellites with $N\ge 30$ particles
within $r_v$ at $t_{mer}$, in order to reduce the importance of
numerical effects on the results. The complete set of nine simulations
yields 461 satellites satisfying this constraint.

\section{Tidal forces and dynamical friction}\label{sec:dftf}

In this section we review qualitatively the physical processes 
dissolving the substructure.

Dynamical friction drives satellites towards energy equipartition
with the smooth distribution of the cluster particles. 
Satellites are more massive than single particles, hence they are
slowed down, their orbits shrink and become more circular
as the satellites lose energy and angular momentum. 
Eventually the satellites are deposited in the center of the cluster 
potential well.

Two different types of tidal forces must be considered during the
infall of satellites onto a cluster: (1) global tides caused by the
interaction with the main halo, and (2) tides caused by encounters
between satellites.  Global tides unbind mass from the satellite in
favour of the deeper potential well of the main halo. For almost
circular orbits this effect is relatively mild, and only the mass in
the outskirts of the halo is stripped off.  Conversely, satellites on
very eccentric orbits pass through the cluster core, and are
completely evaporated by tidal shocks.  Besides global tides, close
encounters with other satellites convert orbital energy into internal
energy, causing collisional stripping and eventually disruption.

Dynamical friction and tides are in action at the same time; however
they have somewhat different effects on a satellite.  Dynamical
friction mostly influences the orbit of a satellite, by always
reducing the orbital energy and angular momentum, making satellites
more susceptible to global tides.  Close encounters between satellites
within the cluster change both the satellite orbit and its internal
structure. Orbits are modified randomly, with no predictable net
effect, and a satellite may either lose part of its mass or even
capture mass from the perturber.  Global tides mainly affect the
internal structure of satellites.

Figure~\ref{fig:part} shows three examples of mergers which roughly
correspond to the extreme behaviors found in the simulations.  They
belong to the same simulation and correspond to three satellites which
merged with the main cluster at the same redshift $z=0.9$.  In the
first row the satellite and the main halo have mass ratio $0.5$.  For
massive mergers like this one, dynamical friction is very efficient, and
the satellite is driven to the center of the main halo in a very short
time, almost regardless of the satellite initial orbit.  The accreted
satellite is also heated by the tidal shock caused by the potential of
the main halo and possibly by close encounters with other
satellites. However, due to its relatively large initial size, a
significant part of its mass is confined within its original virial
radius $r_v$.  In the second and third row the mergers have a much
smaller mass ratio, of the order of $0.01$. For this mass ratio,
dynamical friction time is longer than a Hubble time, and tidal 
forces essentially  determine the disruption or survival of a satellite. 
In the second example the satellite experiences a close
encounter with another satellite inside the main cluster, and
evaporation is immediate.  In the third example, the orbit avoids both
the cluster core and collisions with other satellites, allowing a much
longer survival time. Therefore, the disruption or survival
of a satellite depends on the balance between its initial orbit, close
encounters with other haloes, and the efficiency of dynamical
friction.

\section{Evolution of orbits}\label{sec:orb}

\subsection{Orbits of satellites}\label{sec:orbd}

\begin{figure}
\centering
\epsfxsize=\hsize\epsffile{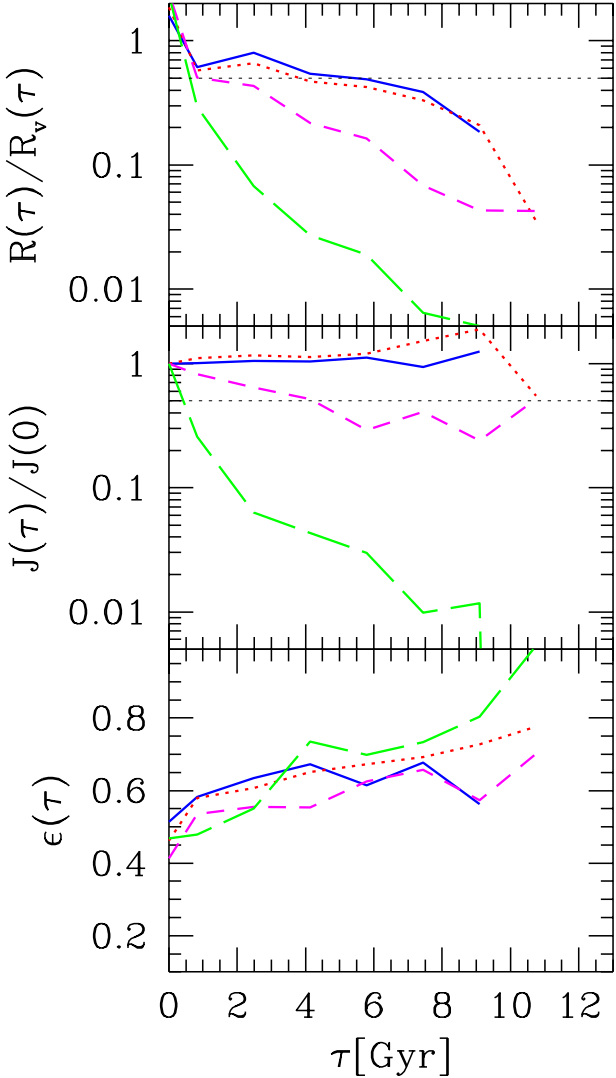}
\caption{Evolution of orbits: distance, angular momentum, and 
circularity. Solid, dotted, short-dashed and long-dashed lines 
are for $m_v/M_v\in (0.00,0.01]$, $(0.01,0.05]$, $(0.05,0.20]$, 
$(0.20,1.00]$, respectively, where $m_v/M_v$ is the ratio between
the satellite mass and the main halo mass at merging time. 
The light dotted line is at half of the initial distance and 
angular momentum.}
\label{fig:orbits}
\end{figure}

In this section we study the systematic orbital decay experienced by
satellites after crossing the virial radius of the main cluster.  We
consider all the time outputs after the merging time $t_{mer}$ and
calculate the main orbital parameters for each $\tau \equiv t -
t_{mer}$. Figure~\ref{fig:orbits} shows the results. To show the
dependence of the orbital decay on the satellite mass, we consider four
subsets, according to the ratio $m_v(t_{mer})/M_v(t_{mer})$ between
the satellite mass $m_v(t_{mer})$ and the mass $M_v(t_{mer})$ of the 
main cluster at merging time.
We then calculate the median value of the parameters over the
satellite distribution for each mass bin and for each $\tau$.

The top panel of Figure~\ref{fig:orbits} displays the evolution 
of the mean distance $R(\tau)$ of a satellite from the cluster center. 
Note that distances at a given time are in units of the virial 
radius of the main halo at that time.
Therefore the median distance decreases both because of dynamical 
friction and because of the increasing size of the main halo.
Interpreting the figure with this {\it caveat}, we observe that the
orbital decay is faster for more massive satellites, as expected
from dynamical friction.
The time required to halve the mean orbital distance is roughly
5 Gyr for satellites with $m_v/M_v \le 0.01$, 4 Gyr for 
$0.01 < m_v/M_v \leq 0.05$, and less than 1 Gyr for $m_v/M_v > 0.05$.
Due to the mass cutoff at 30 particles, no satellite was accreted 
11 Gyr before the final time in the smallest mass bin. Therefore 
the solid curve is less extended than the others. It does not mean 
that the distance drops to zero at $\tau \simeq 11$ Gyr.

The second panel shows the evolution of the orbital angular momentum.
This quantity is almost conserved for satellites with $m_v/M_v \le 0.05$, 
while the halving time is 4 Gyr for $0.05 < m_v/M_v \leq 0.20$, 
and less than 1 Gyr for $m_v/M_v >0.2$.

The bottom panel shows the evolution of the orbital circularity
$\epsilon$, defined as the ratio between the orbital angular 
momentum and the angular momentum of a circular orbit with the same 
energy.
This parameter is a measure of the orbit shape: values of $\epsilon
\sim 0$ correspond to almost radial orbits, while circular orbits
have $\epsilon = 1$. The energy equipartition caused by dynamical 
friction leads to orbital circularization, as indeed observed
in the figure. However, the dependence on the satellite mass
is weaker for circularization than for  
orbital decay and angular momentum loss. Only the most massive satellites 
reach complete circularization in a Hubble time.

\subsection{Comparison with theory}\label{sec:orbt}
Here we compare the actual orbits of the satellites in the $N$-body
simulation with a theoretical prediction based on a simple local
(Chandrasekhar 1943) prescription for dynamical 
friction. The theoretical
orbits were calculated with time-dependent satellite masses.
The orbit of each satellite was integrated in the
time-dependent potential of its parent cluster as measured from the
simulations.  For this purpose, the potential was derived from the
spherically averaged mass profile and its centre of mass was fixed.
For the dynamical friction we applied a frictional force 
\be f_{DF} =
- {\Lambda \rho(t,r) m_{sat}(t)\over s^3} {\bf v}_{sat}
\label{eq:fDF}
\ee 
where $\rho(t,r)$ is the local halo density at the position of the
centre of mass of the satellite, ${\bf v}_{sat}$ is the instantaneous
velocity of the satellite, and $s$ is defined by
\be
s = \max[v_{sat},v_c],
\label{eq:svsat}
\ee
where 
\be
v_c =  (M(R)/R)^{1/2}
\label{eq:vc}
\ee 
is the circular velocity in the halo at the satellite position.
Equation (\ref{eq:fDF}) is a crude approximation to the usual local
Chandrasekhar formula for dynamical friction in an isotropic system
(e{.}g{.} Binney \& Tremaine 1987).  Near the centre of the parent
halo $v_c$ is softened by replacing $R$ with $\sqrt{R^2+\epsilon^2}$,
where $\epsilon\simeq2kpc$. The Coulomb logarithm $\Lambda$ was set to
a constant ($=8$).  We tried two prescriptions for the satellite mass
$m_{sat}$: 1) the `virial' mass $m_v(t)$, more exactly the mass of
satellite particles within the original virial radius at the
identification time; and 2) the self-bound mass $m_b(t)$, defined in
Section \ref{sec:sb}.  The former was found to give the best results
(a similar conclusion was reached by Navarro, Frenk \& White 1995),
and was used for the results described.

\begin{figure}
\centering
\epsfxsize=\hsize\epsffile{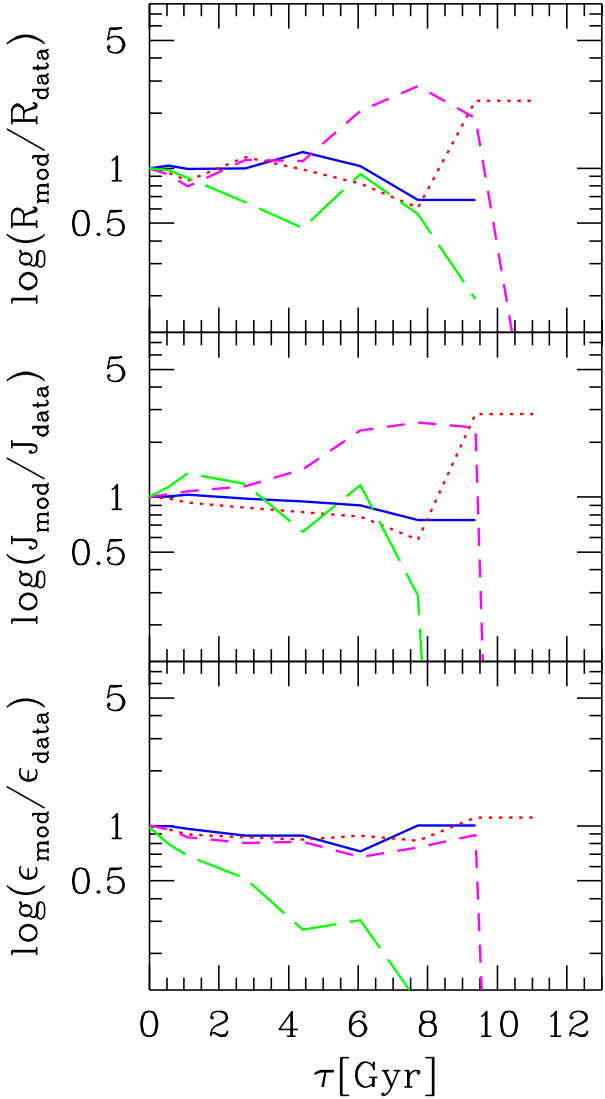}
\caption{Evolution of orbits compared with theory assuming a simple
local prescription for dynamical friction: as in
Figure~\ref{fig:orbits} distance, angular momentum, and circularity.
Each panel shows the ratio of the theoretical prediction to the
quantity measured in the simulation, in each case taking the median
value over all satellites at a given time.  Solid, dotted,
short-dashed and long-dashed lines are for $m_v/M_v\in (0.00,0.01]$,
$(0.01,0.05]$, $(0.05,0.20]$, $(0.20,1.00]$, respectively.}
\label{fig:friction}
\end{figure}

The results are shown in Figure~\ref{fig:friction}.  Here we show the
ratio of the theoretically predicted to the measured quantities in
terms of the median over all satellites in each mass bins (the same
mass bins as defined for Figure~\ref{fig:orbits}).  In all cases the
theoretical prediction is rather good for the two smallest mass bins,
and fails for the large mass satellites as they approach the centre.
This is not surprising since the assumptions behind the simple orbit
model break down when $M(R)$ is comparable with $m_{sat}$.  The most
drastic effect of the breakdown of the assumptions is the plunging of
the theoretical orbits of the large satellites towards the centre of
the parent cluster, with $J(\tau)$ and $\epsilon(\tau)$ tending
rapidly to zero.  Even for these most massive satellites, however, 
the evolution up to $\sim 2$ Gyr of $R(\tau)$ and (to a lesser extent)
$J(\tau)$ is well enough reproduced by the theoretical calculations. 
Thus, the decay times can be accurately predicted.

\section{Survival times}\label{sec:surv}

In this section we consider the global disruption of substructure.
One way to define the {\em survival time} $\tau_{sur}$ of a satellite 
of mass $m$ is through the instantaneous mass loss:
\be
\tau_{sur} = - {m(\tau) \over \dot m(\tau)}
        = - {m(\tau) \over \Delta m(\tau)} \Delta t
\label{eq:tsur}
\ee
where $\Delta m(\tau)=m(\tau+\Delta t)-m(\tau)$. A satellite followed 
inside the cluster for $n$ time outputs will contribute $n-1$ data 
points to the distribution of $\tau_{sur}$. We implicitly assume 
that these points are statistically independent from each other.
Another sensible definition of {\em survival time} $\tau_{sur}$ 
is simply the time taken by a satellite to completely lose its 
initial mass $m_v = m(r_v;t_{mer})$. We will use both definitions
below.

\begin{figure*}
\vbox to13.cm{\rule{0pt}{13.cm}}
\includegraphics{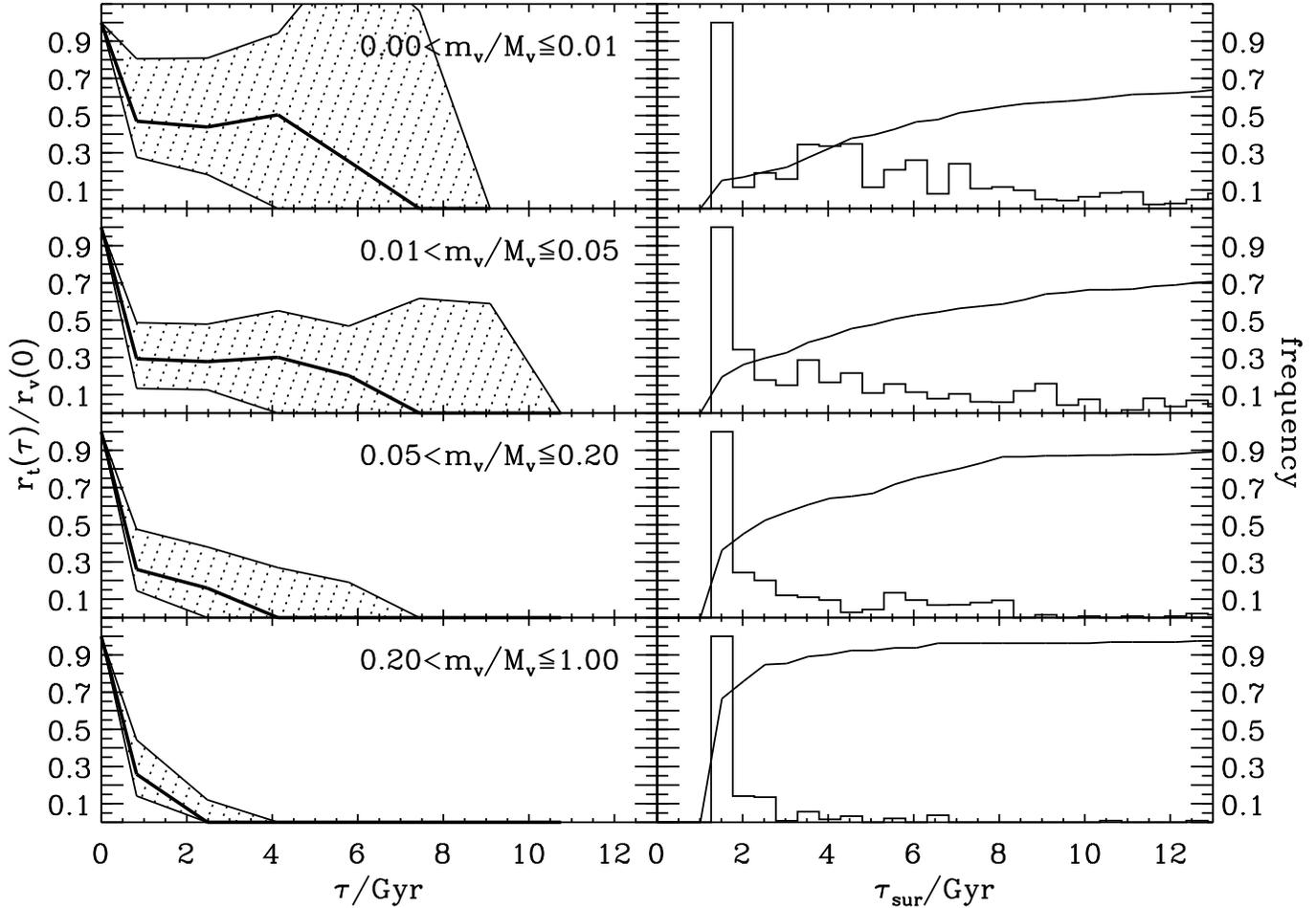}
\caption{Left panels: evolution of tidal radii. Bold solid line is the
median evolution; shaded areas include first and third quartiles. 
Right panels: differential (histograms) and cumulative (solid lines) 
distributions of the survival times defined in Equation
(\ref{eq:tsur}) for the same mass bins of the left panels.
We remind that $m_v$ is the virial mass of satellites at merging time
$t_{mer}$, and $M_v$ is the virial mass of the main cluster progenitor
always at $t_{mer}$.}
\label{fig:tidal}
\end{figure*}

In this section we will measure the mass of a satellite in two
different ways: (1) the mass $m_t(\tau) = m(r_t;\tau)$ within the
satellite {\em tidal radius} $r_t$, (Section
\ref{sec:tidal}), and (2) the satellite self-bound mass $m_{sb}(\tau)$,
(Section \ref{sec:sb}).  Unlike $m_t(\tau)$, the self-bound
mass $m_{sb}(\tau)$ contains information on both the positions and
velocities of the particles.  For both mass measures, we only consider
the particles within the satellite's virial radius $r_v$ at $t_{mer}$.

\subsection{Tidal radius}\label{sec:tidal}

Consider a satellite in a circular orbit at distance $R$
from the cluster center. The orbiting satellite is tidally truncated
at some radius $r_t$, loosely speaking where the differential tidal
force of the cluster is equal to the gravitational attraction of the
satellite: 
\be dF_{tid}(R) \equiv {\partial F \over \partial R}dR =
{Gm \over r_t^2}.
\label{eq:dFtid}
\ee
Assuming that the satellite mass $m$ and its radius $r_t$ are 
negligible compared with the cluster mass and with their relative 
distance $R$, i.e. $m \ll M(R)$, and $r_t=dR \ll R$, we readily obtain
\be
r_t = R \left[{m \over (2 - \partial\ln M/\partial\ln R)M(R)}\right]^{1/3}.
\label{eq:trad}
\ee
Therefore the tidal radius is such that the mean density of the
satellite within $r_t$ is of the order of the mean density of the main 
halo within $R$. This definition  captures the essence of the natural
definition of $r_t$, 
defined as the distance of the center of mass of the satellite from the
saddle point of the potential of the total system.

For non circular orbits, the most common situation in real life, 
the satellite tidal radius is rather ill-defined (e.g. King 1962; 
Binney and Tremaine 1987; Mamon 1987; Mamon 1993). In this case, one 
usually keeps Equation~(\ref{eq:trad}), taking for $R$ the pericentric
distance of the satellite orbit. We adopt this procedure. 
At each time $\tau$, $r_t$ is the solution of the equation 
$\bar{\rho}_{sat}(r_{t}; \tau) =  \bar{\rho}_{main}(r_p; \tau - \Delta t)$, 
where $\bar{\rho}_{sat}(r_{t}; \tau)$ is the satellite mean density at
$r_t$, and $\bar{\rho}_{main}(r_p; \tau - \Delta t)$ is the mean
density of the main halo at the pericenter $r_p$ of the orbit the 
satellite had at the previous time output. 
We compute density profiles for the satellites at each time
considering only the particles within $r_v$ at $t_{mer}$.

We then define the survived mass of a satellite as the mass 
$m(r_t;\tau)$ within its tidal radius. With this mass,
Equation~(\ref{eq:tsur}) provides our first definition of survival
time $\tau_{sur}$ of a satellite.  Note that the tidal radius is
usually associated with the effect of global tides on a static
satellite profile.  We use the {\em actual} density profile of the
satellite, which changes in time accordingly to all kind of
interactions experienced by the satellite.  Therefore, the
corresponding survival times are a measure of the times taken by a
satellite to be destroyed by all processes together: i.e. cluster
tides, close encounters and dynamical friction.

Figure~\ref{fig:tidal} shows the evolution of tidal radii $r_t$ and 
the distribution of survival times for the accreted satellites. 
We bin satellites into four mass bins as in Figure~\ref{fig:orbits}.
As one could expect, tidal radii of small satellites vanish more
slowly than tidal radii of massive satellites.
Notice that in all mass bins $r_t$ drops to 30 - 50 per cent of its
initial value as soon as the satellite enters the main cluster and
finds itself embedded in the denser environment. Because of the 
sudden change in the surrounding density, particles with large 
kinetic energy escape the satellite. The dense core survives, and 
afterwards, $r_t$ declines to zero more gently, with an
intermediate constant phase for the less massive satellites.

The corresponding survival times are displayed in the right panels. 
The distributions are plotted up to a Hubble time, but they
have a long tail further on the right, as indicated by the cumulative
distributions. Median survival times for the four mass bins are,
from low to high mass, $\tau_{sur} = 7$ Gyr, 5.5 Gyr, 2.4 Gyr and less 
than 1.5 Gyr. The fraction of survival times above 13 Gyr is
36 per cent, 30 per cent, 10 per cent and 3 per cent from low to high mass.

\subsection{Self-bound mass}\label{sec:sb}

Another measure of the mass associated with substructure is 
the fraction $m_{sb}(\tau)/m_v(0)$ of the initial satellite 
mass which remains gravitationally self-bound. The {\em self-bound}
mass of a satellite was defined as follows. 
\begin{enumerate}
\item
At any given time we consider only the particles which 
composed the satellite at its identification time.
\item
We estimate the total energy of each particle summing the 
potential energy due to the distribution of these particles
and the kinetic energy calculated in a reference frame moving 
with the average velocity of all particles within the original 
virial radius $r_v$ of the satellite; the reference frame is 
centered on the position found by the moving center technique 
(see Tormen et al. 1997 for details).
\item
We remove all particles with positive total energy, and 
calculate a new center of mass and average velocity 
for the distribution of the remaining particles.
\item
we calculate new total energies using the new set of particles
and the new center of mass and velocity.
\end{enumerate}
We iterate the last two steps until the number of particles with
negative energy is constant. This final value
gives the self-bound mass of the satellite $m_{sb}(\tau)$.

\begin{figure*}
\vbox to13.cm{\rule{0pt}{13.cm}}
\includegraphics{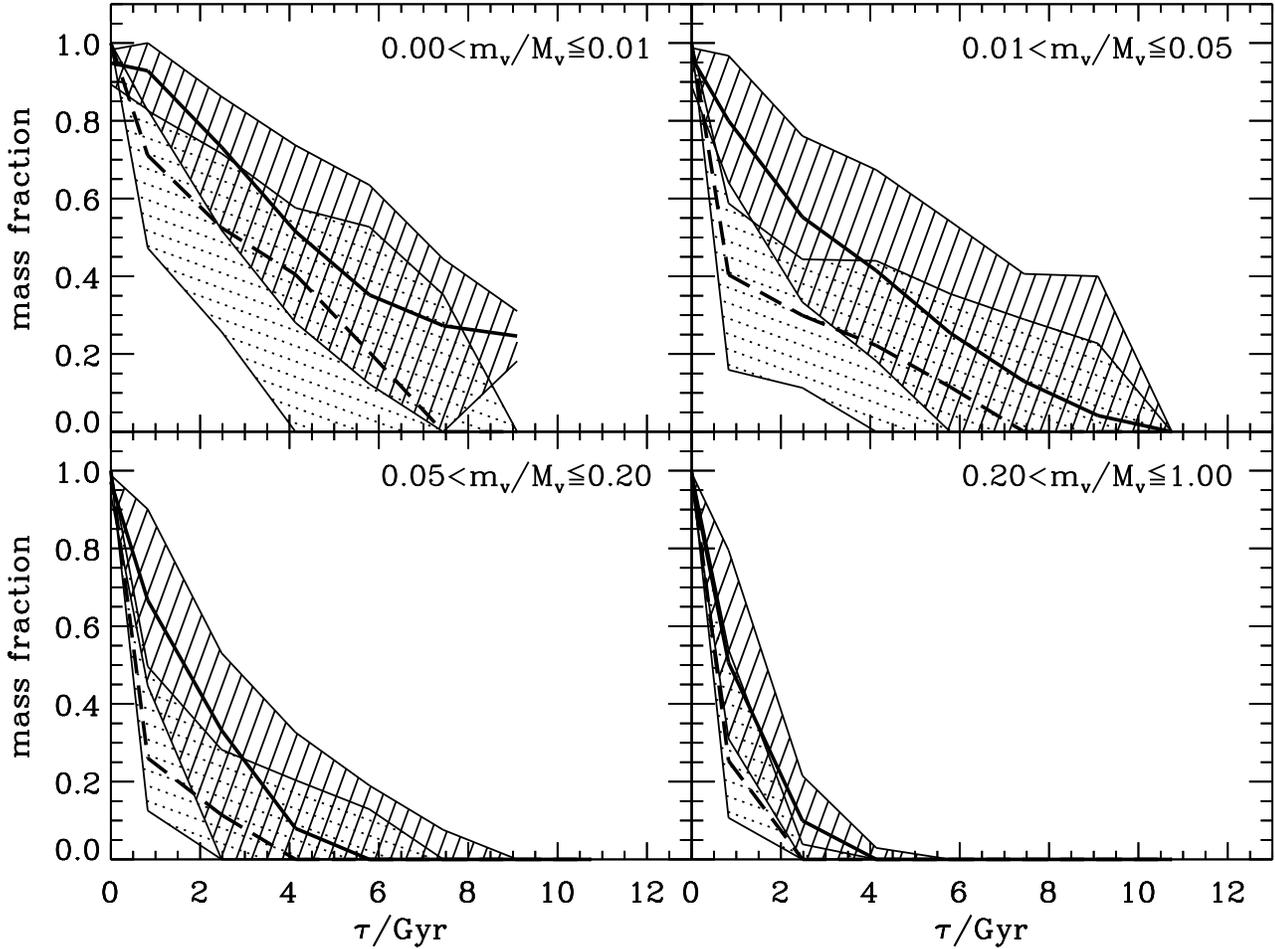}
\caption{Survived mass fractions after merging. Solid and dashed 
lines are the median evolutions of the self-bound mass and the mass 
within the tidal radius, respectively. Shaded areas include first 
and third quartiles. Mass bins are as in Figure~\ref{fig:tidal}.}
\label{fig:mloss}
\end{figure*}

In Figure~\ref{fig:mloss} we compare the evolution of the self-bound 
mass with the evolution of the mass within the tidal radius. 
Bold lines show median values, while shaded areas show the first and 
third quartile range of the distribution at each time $\tau$.
The trend of the two estimates is similar, but the mass within $r_t$ 
is always smaller than the self-bound mass.

It is not surprising that the two definitions do not agree. In fact,
they use rather different criteria to identify the satellites. The
mass within the tidal radius is defined by the spatial distribution of
particles alone, Eq.~(\ref{eq:trad}); the self-bound mass also uses
information coming from the particle velocities.  Because particles in
the outskirt of a satellite will take roughly a free-fall time to be
accelerated to the cluster velocity dispersion, they will stick to the
satellite for some time after they fall outside the tidal radius, and
will be counted in the self-bound mass.  Thus, the estimate
based on tidal radii is actually a conservatively small measure of the
mass associated with the satellite.

We use the evolution of the mass within $r_t$ as our second
estimate of survival times. Figure~\ref{fig:mloss} shows that 
{\em median} survival times, defined by the condition:
$m(r_t; \tau_{sur}) = 0$, are $\tau_{sur} = 7.5,$ $7.5$, $4$ and $2.5$ Gyr
from small to large masses.
These survival times are slightly larger than the survival times
derived with Equation~(\ref{eq:tsur}). However, they are 
consistent with those survival times, because both distributions have 
large scatter.

The longer survival time of small satellites is due to the combination
of two effects: (1) smaller satellites are more compact (Tormen 1997;
see also Section~\ref{sec:instr}); (2) dynamical friction is less
effective on smaller satellites; in fact, smaller satellites have 
larger distances from the cluster center (Figure~\ref{fig:orbits}), 
and suffer a weaker global tide.

These results suggest that the high force and mass resolution of our 
simulations overcome, at least partially, the overmerging problem which 
is common to dissipationless $N$-body simulations (Klypin et al. 1997). 
Moreover, survival of galaxy size haloes do not necessarily need 
dissipative simulations as usually believed (e.g. Summers et al. 1995).
For example, satellites with mass ratio below $0.01$, roughly 
corresponding to galaxy size haloes falling into a forming cluster, 
have median survival time 7-$7.5$ Gyr. 
Thus, half of the satellites of this size, accreted by a cluster 
at $z \simeq 0.6 - 0.8$, are still ``safely'' orbiting within 
the cluster potential at $z=0$. One example is the third satellite 
in Figure~\ref{fig:part}.
Defining survival times based on self-bound masses would result in
even longer survival.

Finally, we find that survival times do not depend on the initial 
orbit of satellites. Radial orbits drive satellites through the 
cluster core where tidal forces dissolve them. However, radial 
orbits are rare (Tormen 1997), and our statistics is too poor 
to investigate this issue satisfactorily.

\section{Encounters between satellites}\label{sec:coll}

In this section we investigate the importance of close encounters
between satellites orbiting within the parent cluster. High speed
close encounters have been advocated as a major mechanism for the
morphological evolution of galaxies in clusters (Moore et al. 1996).
The mass loss in satellite-satellite encounters depends on different
parameters.  Here we restrict our analysis to two quantities: the
relative distance $b$ of the encounter, and the ratio $\gamma$
between the relative tangential velocity $dv_{tg}$ of the two satellites and the
internal velocity dispersion $\sigma_v$ of the perturber. 
Closer (smaller $b$) and slower (smaller
$\gamma$) encounters are more effective in disrupting the
colliding satellites.

\begin{figure}
\centering
\epsfxsize=8.5cm\epsffile{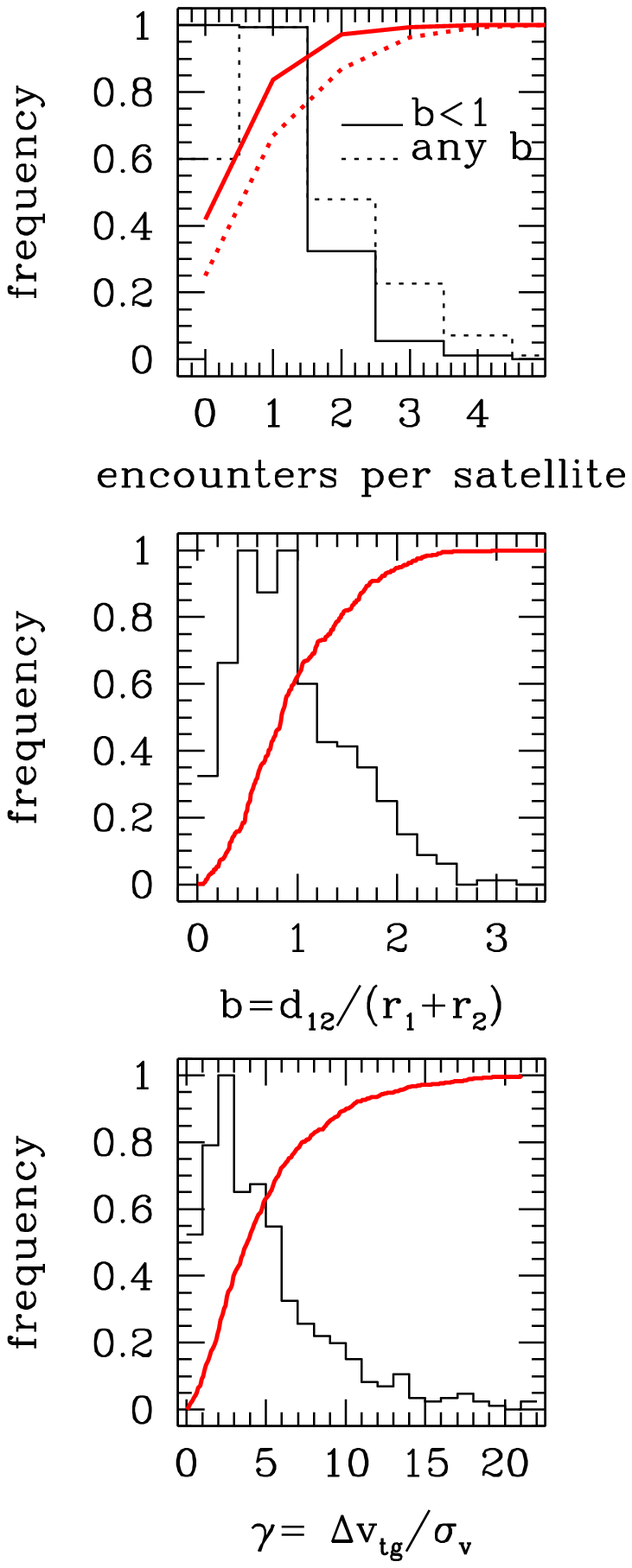}
\caption{Parameters in satellite-satellite encounters.
In each panel the thin and thick histograms indicate the differential 
and cumulative distribution respectively.
{\bf Top panel:} total number of encounters (dotted) and of penetrating 
encounters (solid) experienced by satellites within the main halo.  
{\bf Middle panel:} Dimensionless relative distance of the encounter.
{\bf Bottom panel:} Relative tangential velocity of the encounter,
in units of the internal velocity dispersion of the perturber.}
\label{fig:coll1}
\end{figure}

Operationally, we look for encounters between satellites in the
following way. Consider a satellite $i$ accreted by the main cluster
at time $t_i$.  The other satellites in the main halo are the
`perturbers' of satellite $i$.  At each time output $t>t_i$, we
compute the distance of satellite $i$ from all perturbers.  The
dimensionless relative distance between satellite $i$ and perturber
$j$ at time $t$ is $b(i,j;t) = d(i,j;t)/[r(i) + r(j)]$, where
$d(i,j;t)$ is the distance between the two satellites at that time,
and $r(i),r(j)$ are the virial radii of the satellites at their
respective merging times. The dimensionless relative velocity of the
encounter $\gamma(i,j;t)=dv_{tg}(i,j;t)/\sigma_v(j)$ is similarly defined. The
fraction of self-bound mass lost by the satellite between the time of
the encounter and the next time output is
\be 
{ \Delta m(i) \over m(i)} \equiv {m_{sb}(i;t) - m_{sb}(i;t +
\Delta t) \over m_{sb}(i;t)}.  
\ee
We then consider the distribution of these quantities for all
satellites, at all times $t>t_{mer}$. With the present definitions, a
satellite may provide more than one data point to the distributions.
We make the assumption that each satellites provides statistically
independent data points in different time outputs, as was assumed in
Section~\ref{sec:surv}. We restrict our search to:
\begin{enumerate}
\item
pairs of satellites retaining a non negligible fraction of their original 
mass at the time of the encounter: if $f_{sb}(t) = m_{sb}(t)/m_v(t_{mer})$
is the fraction of the initial satellite mass which remains self-bound
at time $t$, we require that both $f_{sb}(i;t)$ and $f_{sb}(j;t) > 0.2$.
\item
satellites found within the virial radius of the main halo:
$R(\tau)/R_v(t)<1$.
\end{enumerate}
The first requirement excludes fake encounters between haloes which 
have already largely dissolved.
The second requirement excludes time outputs when a satellites has 
left the main halo after a first identification.
We also exclude satellites identified at the penultimate time output of
the simulation, because such satellites are found within the cluster in
just one output (the last one), and so we cannot calculate their
mass loss.

The top panel of Figure~\ref{fig:coll1} shows the differential and
cumulative distributions of the number of encounters per satellite. We
asked each satellite: ``how many encounters did you experience which
satisfy the requirements listed above?''  The answer is illustrated by
the dotted curves, which refer to any value of $b$.  We also asked:
``how many encounters with $b<1$ (referred to as {\em penetrating
encounters}) did you experience which satisfy the same
requirements?''. The answer is given by the solid curves.  The dotted
cumulative distribution shows that 25 per cent of the satellites
accreted by the cluster have no encounter at all that satisfy the
conditions listed above, while the remaining 75 per cent have 1 to 4
encounters.  More interesting, the solid cumulative distribution shows
that almost 60 per cent of all satellites have at least one
penetrating encounter.  We repeated the same census for satellites
with $m_v/M_v <0.01$, representing galaxy-size haloes merging with a
forming cluster. Almost 85 per cent of these haloes have at least one
encounter, and 55 per cent have at least one encounter with $b<1$.  In
the other two panels we show the differential and cumulative
distributions of relative distance and relative velocity for
satellites having at least one encounter.  The central panel
illustrates that penetrating encounters are very common within the
cluster: over 60 per cent of the encounters have $b<1$.  The bottom
panels shows that the relative velocity of the encounter is generally
much larger than the internal velocity dispersion of the
satellites. In fact, the satellites orbit inside the parent cluster at
a speed of the order of the cluster velocity dispersion, which is much
larger than the internal velocity dispersions of the satellites
themselves. As a consequence, only a few per cent of the encounters
have $\gamma$ of order unity.  This high relative speed
justifies the approximation of ignoring mass capture during the
encounters, which we implicitly make when we assigning particles to a
satellite.

\begin{figure}
\centering
\epsfxsize=\hsize\epsffile{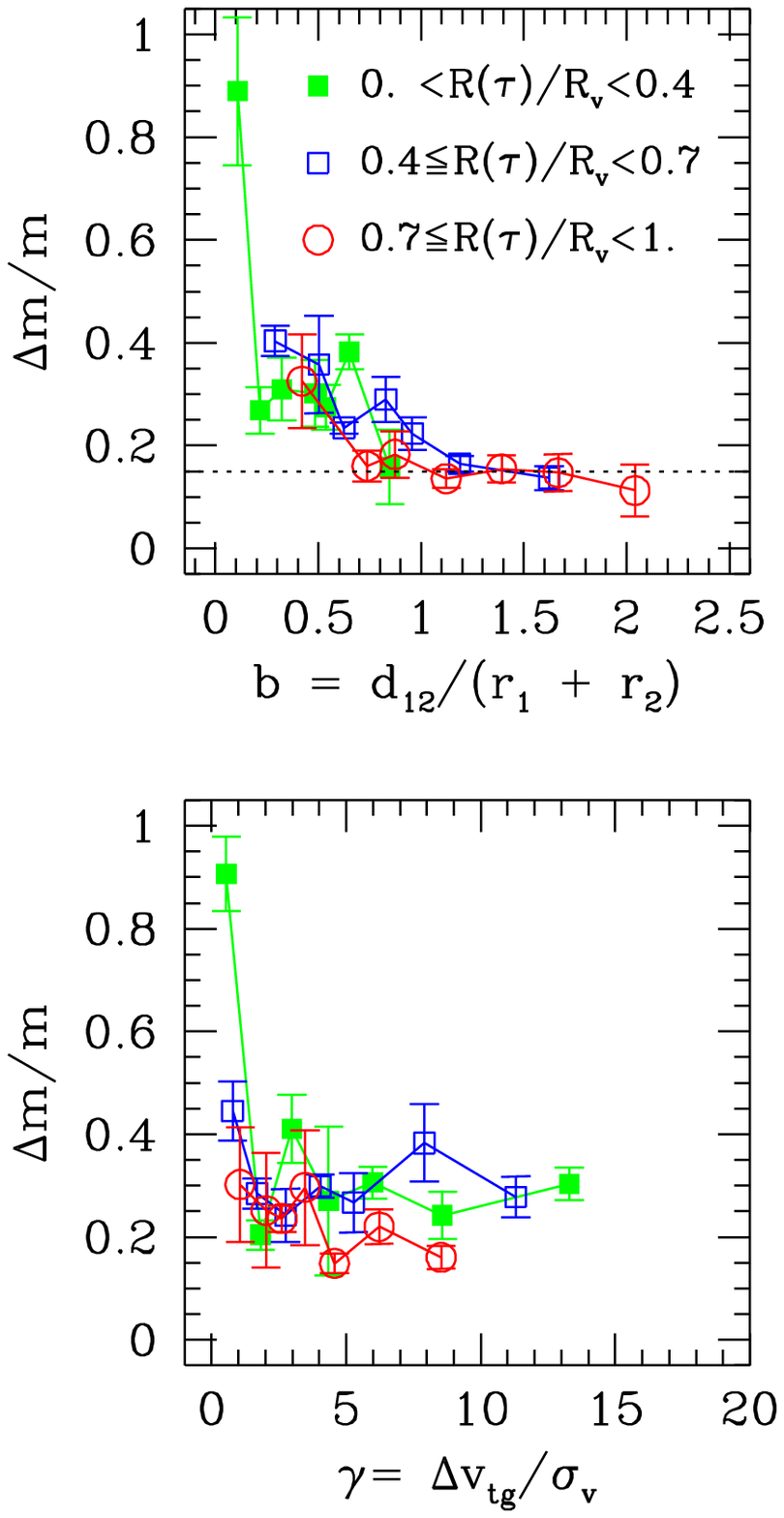}
\caption{{\bf Top panel:} median mass loss versus relative distance of
the encounters $b$. Different symbols refer to encounters occurring at 
different distances from the center of the main halo: 
$0 < R(\tau)/R_v < 0.4$ (solid squares), $0.4 \le R(\tau)/R_v < 0.7$ 
(open squares) and $0.7 \le R(\tau)/R_v <1$ (open circles). 
In each curve the data are binned so that each point represents the 
same fraction of the total encounters. 
Error bars are $\pm 1 \sigma$ of the median.
{\bf Bottom panel:} median mass loss versus relative velocity of the
encounter. Symbols and data binning as in the top panel.}
\label{fig:coll2}
\end{figure}

To disentangle the mass loss due to low speed, close encounters,
from the mass loss due to the tidal forces of the main halo, 
in Figure~\ref{fig:coll2} we consider mass loss versus $b$ (top panel)
and mass loss versus relative velocity (bottom panel), for encounters
occurring at different distances $R(\tau)/R_v$ from the center of 
the main halo.
The top panel shows that, for any fixed $b \ga 0.2$, the mass 
loss is roughly independent of this distance: encounters in the 
central region of the main halo ($0 < R(\tau)/R_v < 0.4$: solid 
squares) and those taking place in outer regions 
($0.4 \le R(\tau)/R_v < 0.7$: open squares, and 
$0.7 \le R(\tau)/R_v <1$: open circles) cause the same mass loss,
within the statistical noise.

Notice the different range of $b$ covered by the curves: the solid
symbols are concentrated at $b<1$, while encounters happening at
larger distances extend further to the right. In fact, the 
number density of satellites grows towards the center of the main 
halo, increasing the chance of close encounters.
The solid square at $\Delta m/m \simeq 0.9$ corresponds to
head-on encounters in the inner part of the main halo, which lead
to total mass disruption.

For $b\ga 1$ the curves are consistent with a roughly constant
mass loss: $\Delta m/m \simeq 0.15$. If we interpret this as
the average mass loss due to global tides only, then the extra 
mass loss is due to the encounter.
The fact that mass loss increases only for penetrating encounters
($b<1$) is consistent with this explanation.

The lower panel shows that the mass loss is almost independent of
the value of the relative velocity $\gamma$, and of
the distance $R(\tau)/R_v$ from the center of the main halo. 
The exception is again the leftmost solid square, which corresponds
to slow encounters in the core of the main halo. 
Encounters in the core of the main halo
are both the closest and the slowest, hence the most disruptive.
In fact, (1) the velocity
dispersion and circular velocity of the main haloes decrease towards 
the center for $r/R_v \la 0.3$ (as shown by the radial profiles in 
Tormen et al. 1997), and (2) the number density of satellites
increases towards the center.

On the whole, Figure~\ref{fig:coll2} indicates that the mass loss during
encounters between satellites depends mainly on the relative distance 
of the encounter, $b$, and very little on the relative velocity, because 
slow encounters are rare, while close encounters are frequent.
Therefore, in the rest of this section we will concentrate our study
on the $b$ parameter.

\begin{figure}
\centering
\epsfxsize=\hsize\epsffile{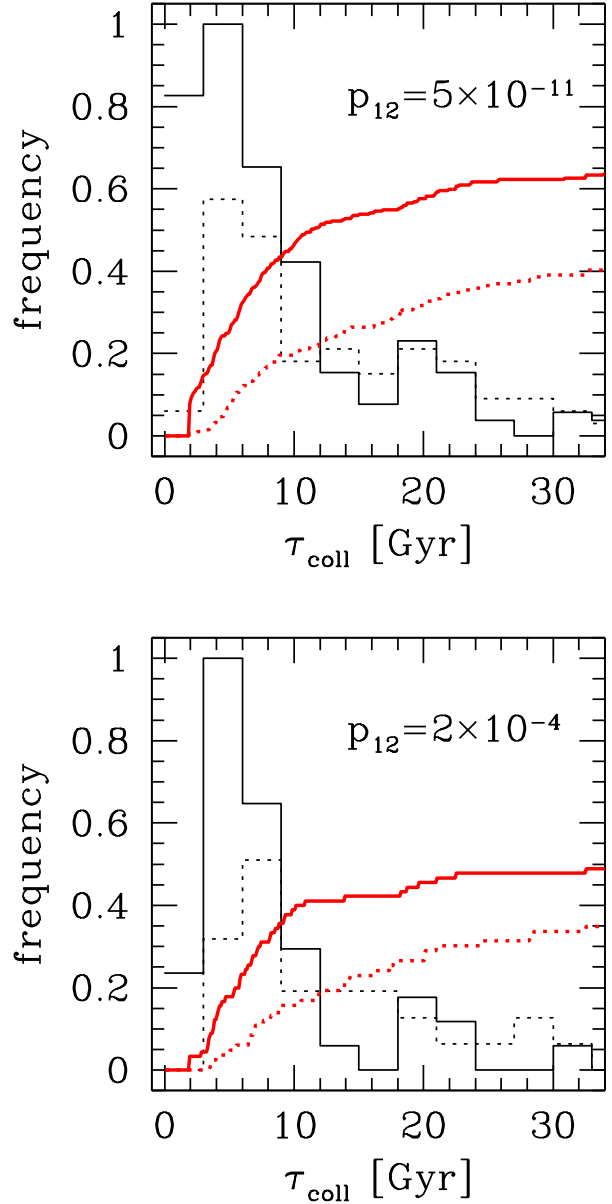}
\caption{Differential (thin histograms) and cumulative (thick curves)
distributions for the disruption time $\tau_{coll}$ from
Eq.~(\ref{eq:tcoll}) due to close encounters between satellites.
{\bf Top panel:} Distribution for satellites of any mass. Solid and 
dotted lines are for $b\in [0,1)$, $[1,\infty)$ respectively.
{\bf Bottom panel:} Same as in top panel, for satellites in the mass 
range $m_v/M_v <0.01$.
Values of $p_{12}$ in each panel are the Kolmogorov-Smirnov 
significance levels for the difference between the solid and 
dotted distribution.}
\label{fig:coll3}
\end{figure}

Since we found that a constant mass loss $\Delta m/m = 0.15$ is a
reasonable zeroth order description of the effect of global tides,
we can estimate the mass loss due only to collisions by subtracting 
this value from that originally measured:
\be
\left({\Delta m \over m}\right)^\prime \equiv {\Delta m \over m} - 0.15.
\label{eq:mcoll}
\ee
The distribution of this quantity is equivalent to the disruption
time associated with the encounters:
\be
\tau_{coll} = -{m(\tau) \over \dot m(\tau)}
             = -\left({m(\tau) \over \Delta m(\tau)}\right)^\prime \Delta t.
\label{eq:tcoll}
\ee

The top panel of Figure~\ref{fig:coll3} shows the differential and 
cumulative distributions of $\tau_{coll}$ for penetrating encounters 
($b<1$: solid lines) and for encounters with $b\ge 1$ (dotted lines). 
For this distribution we used all the data of Figure~\ref{fig:coll2}, 
excluding those contributing to the leftmost solid symbol. 
The median disruption time for $b<1$ is
$\tau_{coll} \simeq 11$ Gyr, corresponding to a median mass loss 
of $(\Delta m/m)^\prime \simeq 0.15$. The percentage of satellites
dissolved by collisions over a Hubble time is 52 per cent.
On the other hand, encounters with $b\ge 1$ have negligible mass loss,
and median $\tau_{coll} \gg t_{Hubble}$, a result consistent with the 
data shown in the top panel of Figure~\ref{fig:coll2}. 

The bottom panel of Figure~\ref{fig:coll3} shows the same statistic for satellites
with $m_v/M_v <0.01$, which correspond to galaxy-size haloes falling
onto a cluster. In this case mass losses due to penetrating encounters
are smaller, with median value $\Delta m/m = 0.05$, corresponding to a
median disruption time $t\simeq 30$ Gyr, with 42 per cent of
satellites being dissolved over a Hubble time.

We can use the data in the first panel of Figure~\ref{fig:coll1}, 
on the frequency of penetrating encounters in satellites, to 
quantify the statistical significance of encounters. 
Since 60 per cent of all satellites have one or more penetrating 
encounters, from the top panel of Figure~\ref{fig:coll3} we can say
that at least 30 per cent of {\em all} satellites are dissolved in 
11 Gyr or less by penetrating encounters alone, and a comparable
fraction is dissolved over a Hubble time.
Similarly, 23 per cent of satellites with $m_v/M_v <0.01$ are 
dissolved by penetrating encounters in less than a Hubble time.
These numbers are an estimate of mass loss and 
disruption time due only to penetrating encounters. Total mass 
losses and total survival times are those presented in 
Section~\ref{sec:surv}.

Finally, we found that these distributions are fairly robust to 
variations of parameters. We obtain essentially the same results
if we change the requirements on the fraction of initial satellite 
mass which must be self-bound at the time of the encounter,
or if we only consider encounters at larger distances $R(\tau)/R_v$ 
from the center of the main halo.

\section{Internal structure: mass and velocity profiles}\label{sec:instr}

\begin{figure*}
\vbox to13.cm{\rule{0pt}{13.cm}}
\includegraphics{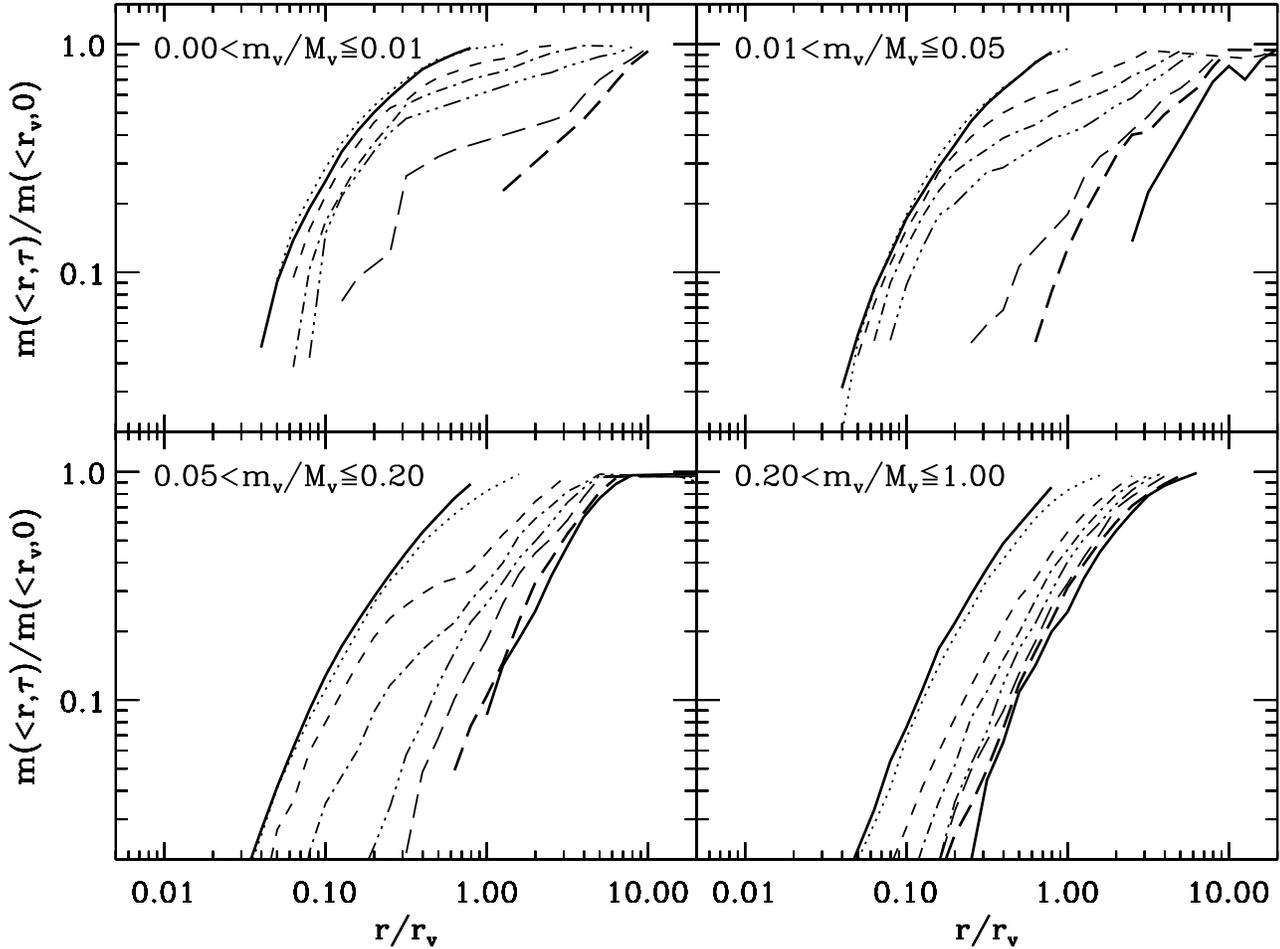}
\caption{Evolution of satellite mass profiles. Lines refer to 
different times. Time sequence is as follows: solid, dotted, 
short-dashed, dot-dashed, dot-dot-dot-dashed, long-dashed, bold 
long-dashed, solid. Note how the mass profile of the small satellites 
(top panels) is severely disturbed at later times (long-dashed, bold 
long-dashed, solid curves). Massive satellites (bottom right panel) 
merely expand self-similarly.}
\label{fig:mprof}
\end{figure*}

\begin{figure*}
\vbox to13.cm{\rule{0pt}{13.cm}}
\includegraphics{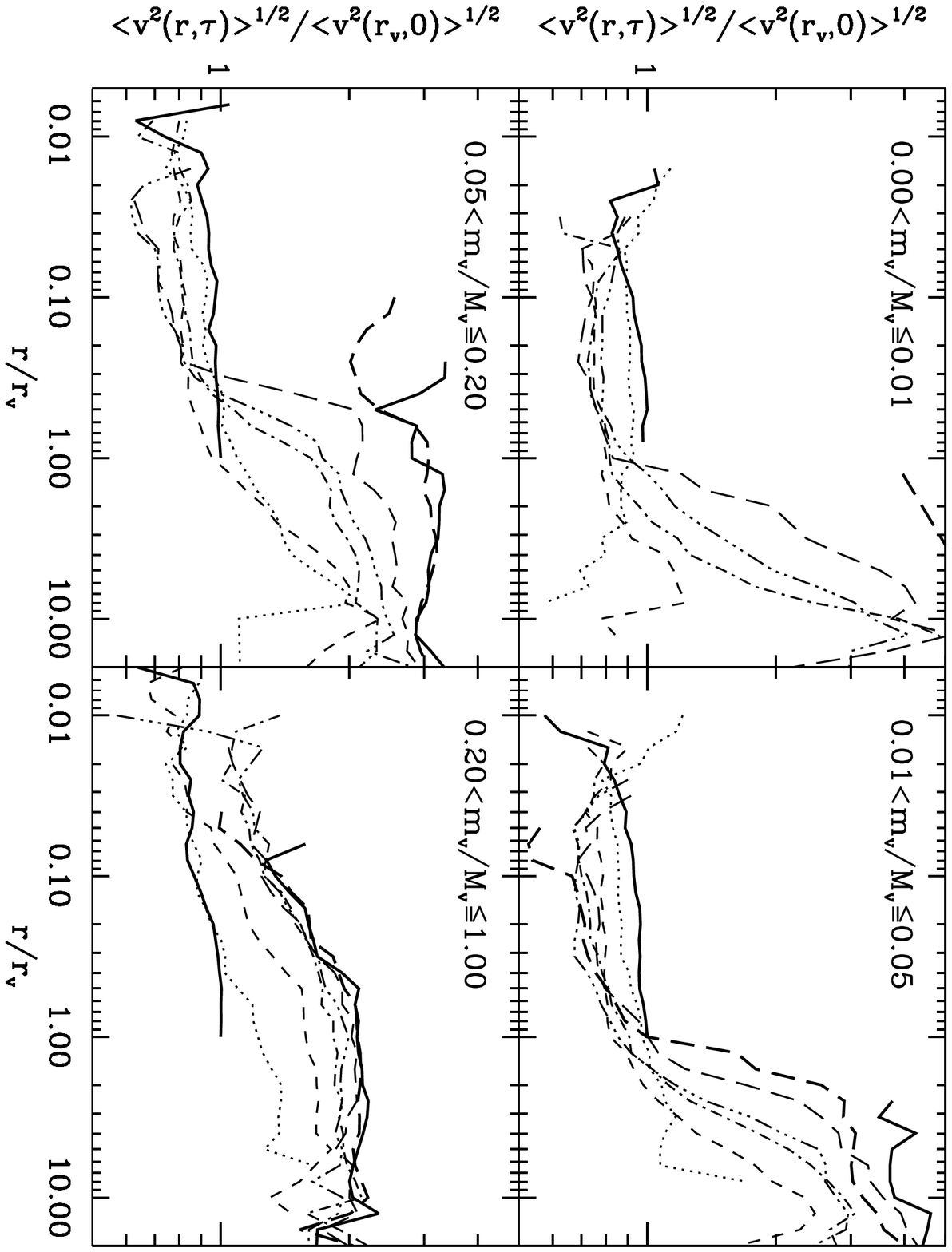}
\caption{Same as Figure~\ref{fig:mprof} for the \rms velocity profiles.
In the top left panel, note how particles first stream out of the 
satellite (dotted curve) and they increasingly thermalize with the 
main halo \rms velocity later on (short-dashed, dot-dashed, 
dot-dot-dot-dashed, long-dashed, bold long-dashed curves). 
The \rms velocity variation is abrupt, indicating that 
satellites are physical entities distinguishable from the main halo.
The thermalization process is smoother with increasing satellite mass. 
The central part of the most massive satellites (bottom right
panel) suffer very little cooling. Finally, note that, on average, 
small satellites are completely disrupted at later times (bold 
long-dashed, solid curves).}
\label{fig:vprof}
\end{figure*}

Here, we investigate the evolution of the internal structure of 
satellites. All the particles composing the satellite just before 
the merging time define the density and the velocity  
profile at each time. We can thus investigate
how the internal structure of a satellite changes after 
the merging time, and how its mass is redistributed within 
the main halo.

The internal structure evolution of satellites is different for
small and large mass satellites.
The density of the environment surrounding a satellite suffers an abrupt
change when the satellite enters the main halo. The tidal radius 
drastically decreases (Figure~\ref{fig:tidal}) and particles with enough
kinetic energy escape the satellite. This process leaves low mass 
satellites with particles having a smaller \rms velocity 
(Figure~\ref{fig:vprof}), producing an overall effect of cooling and a 
slight mass concentration (Figure~\ref{fig:mprof}). Escaped particles 
quickly thermalize to the main halo velocity dispersion.

From Figure~\ref{fig:mprof} we see that small satellites are more 
concentrated than massive ones. In fact, half-mass radii for the 
initial profiles are at $r/r_v= 0.2$, $0.28$, $0.35$, $0.41$ from 
small to large masses. 
For this reason, when we consider more massive satellites, the 
equality between the mean density of satellites and that of the 
main halo occurs at lower values of $r/r_v$. Thus, massive 
satellites are heated well within $r_v$, and the cooling effect 
disappears.

Moreover, the initial \rms velocity of massive satellite particles
is closer to the \rms velocity of the main halo; thus, their kinetic 
energy gain is less dramatic. Mass profiles also show that the 
``disruption'' of massive satellites is actually a mere inflating 
(see also Figure~\ref{fig:part}). 
In fact, the mass profiles are simply shifted to larger values of 
$r$, but their shape remains similar. On the other hand, the external 
shells of small satellites are violently stripped off and particles
are redistributed within the main halo.

\section{The fraction of halo mass in substructure}\label{sec:msub}

From our study it appears that only relatively small satellites can 
resist dynamical friction and survive in a cluster for a significant 
time. However, most of the mass forming a cluster comes from a few, 
massive objects (e.g. Tormen 1997). Therefore, it is natural to ask 
what is the fraction of cluster mass which, at any given time, 
is bound to substructure.
The mass $m(r_t; \tau)$ within the tidal radius is a measure of such
fraction. By definition, the mean density within $r_t$ is larger than 
the mean density of the main cluster within the radius corresponding to
the pericentre of the satellite. Therefore, the mass within the tidal
radius always corresponds to an enhancement in the local mean
density.  We refer to this as a `visible' structure.  By this
definition any satellite with $r_t>0$ is visible.

\begin{figure}
\centering
\epsfxsize=\hsize\epsffile{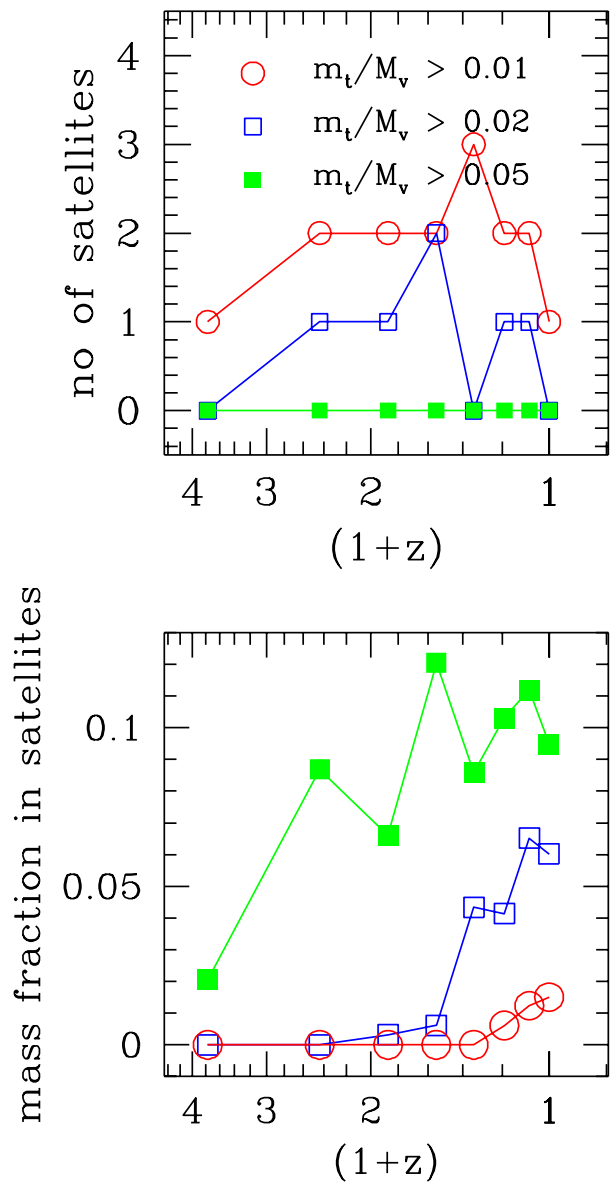}
\caption{Fraction of cluster mass in substructure. {\bf Top panel:}
each curve represents the median (over the cluster sample) number
of satellites found within the main halo, with mass larger than
the threshold indicated in the figure, as a function of redshift.
{\bf Bottom panel:} Median values (over the cluster sample) of 
the fraction of cluster mass associated to satellites, as a function 
of redshift. Open circles, open squares and solid squares refer to
the fraction within a sphere of radius $0.2 R_v$, $0.5 R_v$ 
and $ R_v$ respectively.
Masses are defined using tidal radii, as in Section~\ref{sec:tidal}.}
\label{fig:msub}
\end{figure}

The top panel of Figure~\ref{fig:msub} shows, as a function of
redshift, the median number of satellites within the main halo, and
with mass $m(r_t;t)/M_v(t)$ above a given threshold; $M_v(t)$ is the
virial mass of the main halo at that time.  It shows that, on average,
there are 2 satellites with tidally truncated mass larger than $0.01
M_v$, one with mass larger than $0.02 M_v$, while there is no
satellite larger than $0.05 M_v$.  Note that these mass thresholds
roughly correspond to satellites with radius 20, 27 and 37 per cent 
that of the main halo. 
Such objects should be fairly easily identified by e.g. weak lensing
observation of the dark matter distribution in galaxy clusters 
(see e.g. Geiger \& Schneider 1997).

The mass attached to satellites is shown in the bottom panel. 
The data show that, for most redshifts of interest, substructure 
make up on average 10 per cent of the cluster mass within $R_v$ 
(solid squares). The maximum value measured is 30 per cent.
In the cluster inner regions the fraction of mass in substructure 
lowers dramatically (open squares and open circles refer to spheres 
of radius $0.5 R_v$ and $0.2 R_v$, respectively), due to the stronger 
tidal forces and to the numerical resolution.
This estimate is really a lower limit, as the mass within $r_t$ 
is a conservatively low estimate of the mass of a satellite.
Nevertheless most of the cluster mass is not in substructure, 
but is smoothly distributed inside the cluster.
Therefore the issue of survival of substructure in massive haloes
mainly applies to small ($m_v/M_v \la 0.05$) and compact satellites
rather than to large ($m_v/M_v \ga 0.05$) ones.

\section{Discussion}\label{sec:disc}

\subsection{The extension of dark matter haloes in cluster galaxies}
\label{sec:gsize}

Studies of the kinematics of satellites of field spirals indicate that
these galaxies are embedded in dark halo at least 10 times larger than
the optical radius of the galaxy (Zaritsky \& White 1994; Zaritsky et
al. 1997).  In cluster galaxies, such haloes are thought to be
tidally-truncated by the cluster potential. Unfortunately,
observations of this effect are very difficult. As for spiral
galaxies, they are depleted of their gas by ram pressure in the
intracluster medium soon after they merge, so that rotation curves
cannot be measured at distances significantly larger than the optical
radius (Gunn \& Gott 1972, White et al{.} 1991).  The dark haloes of
ellipticals may in principle be probed by the kinematics of their
globular clusters, but these objects are quite faint, so that HST
observations are needed to study globulars, even in nearby cluster
galaxies. If dark haloes turn out to be small enough, tidal truncation
could be directly observable using weak gravitational lensing (Geiger
\& Schneider 1997) or observing a suitable tracer population to large
radii.

Lacking observational evidence, we can at least use the results of
Section~\ref{sec:tidal} to constrain the expected radius of dark matter
haloes in cluster galaxies. We can associate a galaxy radius to each
dark matter satellite using the dataset of Burstein et al.
(1997). These authors showed that the dynamical properties of most
stellar systems, ranging from globular clusters to galaxies to galaxy
groups and clusters, obey relations similar to the Fundamental Plane
of ellipticals (Dressler et al{.} 1987; Djorgovski \& Davis 1987). 
The ensemble of these planes, termed the
{\em cosmic metaplane}, is an expression of the virial theorem, tilted
by variations in mass-to-light ratio.  In particular, Burstein et
al{.} present data for $\sim 900$ elliptical and spiral galaxies, for
which they derive effective radii $r_e$ (i.e. radii enclosing half of
the total light from the galaxy) and central velocity dispersions
$\sigma_c$.

\begin{figure}
\centering
\epsfxsize=\hsize\epsffile{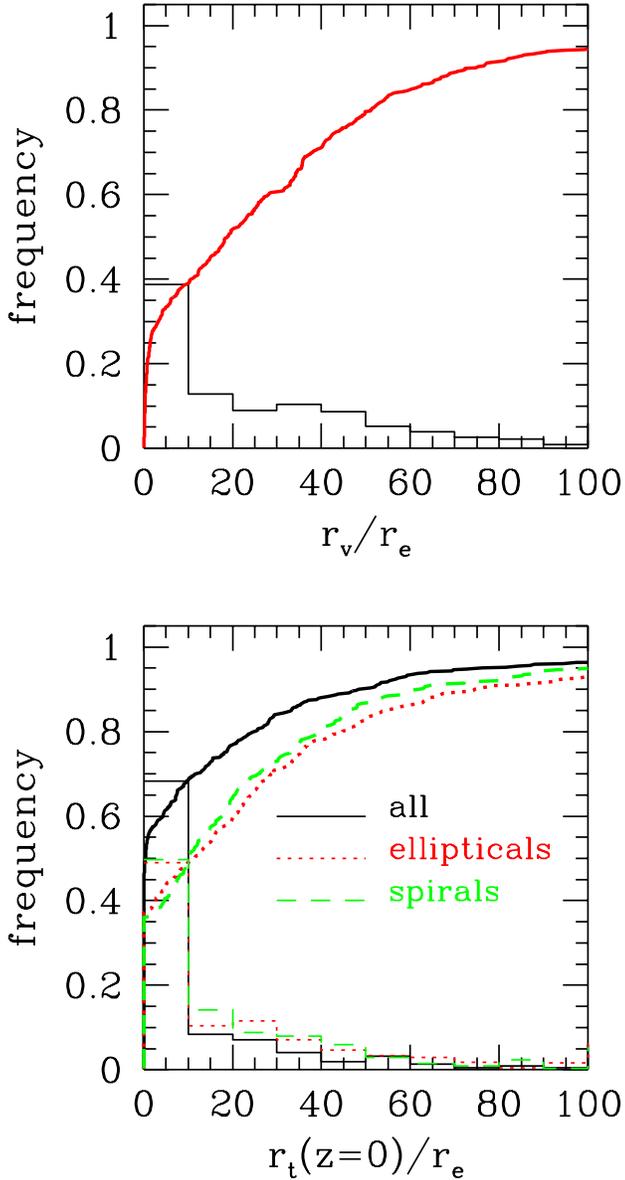}
\caption{Radius of dark matter satellites in units of the effective 
radius $r_e$ of the host-galaxy. {\bf Top panel:} virial radius,
measured at merging time. {\bf Bottom panel:} tidal radius, 
measured at the present time ($z=0$). Details about the choice
of $r_e$ are given in the main text. Thin histograms refer to 
differential distributions, thick curves to cumulative distributions. 
Curves include all the nine cluster simulations.}
\label{fig:gsize}
\end{figure}

Our procedure to assign $r_e$ to each satellite is as follows:
{\em (i)} we consider all satellites infalling onto our simulated 
clusters, and take the one dimensional \rms velocity within the
virial radius, measured at merging time, as a rough estimate of 
$\sigma_c$. {\em (ii)} From the sample of Burstein et al. (1997) we select
all objects having $\sigma_c$ within 10 per cent of the value found 
for the satellite. {\em (iii)} We randomly draw one of these 
objects and assign the corresponding value of $r_e$ to the satellite. 
This prescription can be applied also to a restricted
subsample or class of observations, e.g. only the spirals or
ellipticals in the Burstein et al. database. In such a case, 
satellites with a value of $\sigma_c$ not matched by any of the 
selected observations are excluded, leaving us with satellites 
in a range of $\sigma_c$ appropriate to the choice made.

The result of this exercise is shown in Fig~\ref{fig:gsize}. 
In the top panel we plot the differential (thin histogram) and 
cumulative (thick curve) distribution for the virial radius of 
satellites, in units of $r_e$, with $r_e$ chosen from either 
elliptical or spiral galaxies. This distribution gives an idea 
of the radius of galaxies hosted in dark matter haloes. It shows 
that the median virial radius for merging satellites is rhoughly 
20 times larger than $r_e$.

In the bottom panel we plot similar distributions for the {\em final}
tidal radius of satellites (i.e{.} measured at $z=0$).  The solid
distribution refers to all satellites, with $r_e$ chosen from
galaxies, galaxy groups and galaxy clusters, in order to cover the
all range of satellite masses.  The dotted distribution
refer to satellites associated to $r_e$ chosen from elliptical
galaxies, while the dashed distribution has $r_e$ chosen from spiral
galaxies.  The panel shows that $\approx 35-40$ per cent of
galaxy-sized dark matter satellites ever accreted by a cluster have
been dissolved by the present time, i.e{.} they have $r_{tid} = 0$
(dotted or dashed curves). The median tidal radius for galaxy-sized
satellites is $r_{tid} \simeq 10 r_e$. If we consider all satellites
(i.e. also larger than galaxy-size), half of the satellites are
dissolved by $z=0$ (solid curve).

Strictly speaking, these figures should be regarded as upper limits, 
because an increase in numerical resolution or the presence of a 
dense core of baryonic matter would probably facilitate the survival 
of satellites. The relevance of the latter effect is however not
clear, as dark matter is the dominant source of gravity already at 
the optical radius of the galaxy (e.g. Navarro, Frenk \& White 1996). 
Taking median values as robust estimators of the distributions,
this result on the whole suggests that 50 per cent of galaxies in 
present-day clusters should have a dark matter halo truncated at 
or below $\approx 10 r_e$, a radius close to those radii testable by 
e.g. high precision surface photometry.
Of course, some of these galaxies will have lost a substantial part 
of their dark matter halo and may themselves be disrupted.

\subsection{Interacting galaxies in clusters}

Observations indicate that the fraction of morphologically disturbed 
galaxies or interacting galaxies in clusters typically increases with 
redshift (e.g. Lavery \& Henry 1988; Oemler, Dressler \&
Butcher 1997).
Due to the high relative speed of most encounters, illustrated in 
Figure~\ref{fig:coll1}, mergers between galaxies are expected to be 
rare.
On the other hand, the numerical simulations of Moore et al. (1996) 
have shown that repeated high speed encounters between spirals, at
a relative distance less than a few optical radii (e.g. $\la 50$ kpc 
for an $L_*$ galaxy) can seriously modify the galaxy morphology.
This distance is also the typical separation of the interacting 
galaxies found by Lavery \& Henry (1988).

\begin{figure}
\centering
\epsfxsize=\hsize\epsffile{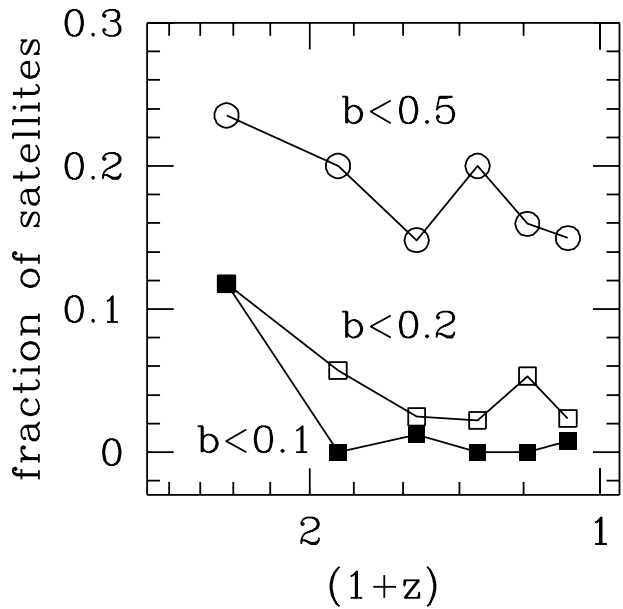}
\caption{Interacting satellites. The panel shows the fraction of
satellites with $m_v/M_v \le 0.01$ experiencing encounters with 
other satellites, as a function of redshift and binned according
to the dimensionless relative distance $b$.
Circles refer to encounters with $b < 0.5$, open and solid squares
to encounters with $b < 0.2$ and $b < 0.1$ respectively.}
\label{fig:coll4}
\end{figure}

We can use the results of the analysis performed in
Sections~\ref{sec:coll} and \ref{sec:gsize} to estimate the fraction of
interacting galaxies in our simulations.  Figure~\ref{fig:coll4} shows
the fraction of galaxy-size satellites undergoing close encounters
with other satellites, as a function of redshift. We show results for
dimensionless relative distances $b<0.5$, $b<0.2$ and $b<0.1$, which
correspond to encounters at distances $<10 r_e$, $<4 r_e$ and $<2 r_e$
if one takes the median value $r_v/r_e \simeq 20$ found in the
previous section as representative. Encounters at such distances
already fall in the class discussed by Moore et al.  We see for
example that $\approx 20$ per cent of these satellites are engaged in
encounters with $b<0.5$, while $\approx 5$ per cent are having
encounters with $b<0.2$, and $1-2$ per cent encounters with
$b<0.1$. These figures are consistent with the $10-20$ per cent of
merging and interacting galaxies observed by Oemler et al.  (1997) in
four clusters at $z \simeq 0.4$. The self-similarity of our
simulations should ensure that such effects do not exhibit any trend
with redshift, as it is indeed the case at least for $z < 1$.

Our results support the claim that close encounters between galaxies 
in clusters are fairly frequent. 
The {\em rate} of encounters produced by our simulations is lower 
than the one invoked by Moore et al{.} for galaxy harrassment, as
most satellites in our simulations have at most one encounters
at $b<1$ (as shown in the top panel of Figure~\ref{fig:coll1}), and 
not several as required by galaxy harrassment.
However, the rate we found should be considered as a lower limit, as 
in real clusters collisions mostly take place in the central, densest
region of the cluster, where we do not have enough resolution to 
preserve the satellite identity. Further investigation with higher
resolution simulations is required to estimate this effect.

\subsection{Comparison with other work}\label{sec:comp}

There exist two different approaches to studying the properties
of substructure within dark matter halos: {\em (i)} we can identify 
satellites before they merge with the main halo and keep track of 
their particle trajectories after merging;
{\em (ii)} we can identify overdensity peaks within the main halo 
at each output time. The latter approach has the advantage of 
identifying physically motivated substructure. 
Moreover, this approach is equivalent to the observational procedure 
when it is applied to two-dimensional projections.
However, this method is not straightforward if we are interested
in the evolution of the substructure properties. In fact, in order 
to check the identity of density peaks identified at different times,
we need to check whether a substantial fraction of the particles of a
density peak at a later time belonged to a density peak at an earlier 
time (Klypin et al. 1997).

In the present work, we have eight output times for each of the nine 
simulations. Therefore, we use the former approach, which is more 
straightforward in keeping track of each satellite. The relatively 
large number of output times is the most important
feature of our analysis. In fact, we can consistently
compare the satellite orbit evolution with
the dynamical friction predictions. Moreover, we can accurately estimate
the evolution of the satellite internal properties
as tidal effects and collisions disrupt them. This information
provides a quantitative definition of the survival time.

Our analysis complements Ghigna et al. analysis.
Ghigna et al. identify overdensities in a single halo 
simulation with higher space and force resolution than ours. 
Thus, they can accurately estimate the satellite density profiles, 
and the spatial distribution of the satellites within the main halo. 
On the other hand, our work has the advantage of a better statistics 
provided by the nine independent clusters and the use of a larger 
number of output times.

\section{Conclusions}\label{sec:conc}

The results of this paper may be summarized as follows:
\begin{enumerate}

\item
The orbital decay of haloes within galaxy clusters is 
consistent with the expectations of dynamical friction. 
Substructure are driven to the center of the main halo in less than
a Hubble time if their initial mass is larger than one per cent
of the mass of the main cluster.

\item
Substructure retain their identity for a significant fraction of the 
Hubble time if their initial mass is smaller than 5 per cent of the 
main cluster mass. 
Median survival times, based on the mass within the tidal radius, 
are in the range [7,7.5] Gyr, [5.5,7.5] Gyr, [2.4,4] Gyr and
[1,2.5] Gyr for mass ratios $m_v/M_v\in (0.00,0.01]$,
$(0.01,0.05]$, $(0.05,0.20]$, $(0.20,1.00]$, respectively.
However, the mass within the tidal radius is conservative, since
survival times are more than 50 per cent longer when we consider 
the satellite self-bound mass.
Smaller satellites have longer survival times for a combination of 
two reasons: {\em a)} they are more compact, and {\em b)} they are 
less influenced by dynamical friction, and avoid the 
cluster core.

\item
Encounters between satellites within the cluster are frequent and 
lead to mass loss comparable to that caused by global tides.
The mass loss is correlated with the relative distance and almost
uncorrelated with the satellite relative velocity. In fact, 
slow encounters are rare, but close encounters are frequent.
Almost 60 per cent of the satellites experience at least one 
penetrating encounter with another satellite before losing 80 
per cent of their initially self-bound mass.
The median mass loss per penetrating encounter is 
$\Delta m/m \simeq 0.15$, corresponding to a median disruption 
time $\tau_{coll} \simeq 11$ Gyr, due only to penetrating encounters. 
The same figures become $\Delta m/m \simeq 0.05$ and
$\tau_{coll} \simeq 30$ Gyr for galaxy-sized satellites.

\item
The evolution of the satellite internal structure depends on the 
satellite mass: smaller satellites easily loose their less bound 
particles and cool in their inner region, while larger satellites 
experience a global heating.

\item
The fraction of cluster mass in tidally-defined substructure is
10 per cent on average within the virial radius, and lower in
the inner parts. It therefore constitutes only a minor fraction 
of the total cluster mass.

\end{enumerate}

The application of our results to galaxies in clusters requires us to
specify a procedure for populating the smaller dark matter satellites
with stellar material.  We discuss an empirical method of assigning
galaxies to haloes in Section \ref{sec:disc}.  The results can be
summarised as follows:

Roughly 50 per cent of galaxies in present-day clusters should have 
a dark matter halo truncated at or below $\approx 10 r_e$.
Some of these galaxies may themselves have been disrupted.

The fraction of satellites undergoing very close encounters,
$b<0.2-0.5$, is similar to the fraction of interacting or merging 
galaxies in clusters at moderate redshift. 
Repeated close encounters, as required by galaxy harrassment, are 
however very rare with the present numerical resolution.

\section*{ACKNOWLEDGEMENTS}
We thank Joerg Colberg, Riccardo Giovanelli, Martha Haynes, 
Julio Navarro and Simon White for useful comments and suggestions.
Financial support for G.T. was provided by an MPA guest postdoctoral
fellowship and by the Training and Mobility of Researchers
European Network ``Galaxy Formation and Evolution''. 
A.D. holds the grant ERBFMBICT-960695 of the Training and
Mobility of Researchers program financed by the European Community.
The simulations were performed at the Institut d'Astrophysique de
Paris, which is gratefully acknowledged.

\end{document}